\documentclass[runningheads]{llncs}
\usepackage{graphicx}
\usepackage[table]{xcolor}
\usepackage{stackengine}
\usepackage{graphicx,subfigure}
\usepackage{amsmath,amssymb}
\usepackage{hyperref}
\usepackage{flushend}

\usepackage{fancyhdr}
\begin{document}
\title{\LARGE \bf Deep Features for Tissue-Fold Detection\\ in Histopathology Images  }

\author{Morteza Babaie$^1$ \and H.R. Tizhoosh$^{1,2}$}
\authorrunning{Babaie \& Tizhoosh}
\institute{$^1$ Kimia Lab, University of Waterloo,  Canada, kimia.uwaterloo.ca\\
$^2$ Vector Institute, Toronto, Canada}

\maketitle              
\begin{abstract}
    
Whole slide imaging (WSI) refers to the digitization of a tissue specimen which enables pathologists to explore high-resolution images on a monitor rather than through a microscope. The formation of tissue folds occur during tissue processing. Their presence may not only cause out-of-focus digitization but can also negatively affect the diagnosis in some cases. In this paper, we have compared five pre-trained convolutional neural networks (CNNs) of different depths as  feature extractors to characterize tissue folds.  We have also explored common classifiers to discriminate folded tissue against the  normal tissue in hematoxylin and eosin (H\&E) stained biopsy samples. In our experiments, we manually select the folded area in roughly 2.5mm $\times$ 2.5mm patches at $20$x magnification level as the training data. The ``DenseNet'' with 201 layers alongside an SVM classifier outperformed all other configurations. Based on the leave-one-out validation strategy, we achieved $96.3\%$ accuracy, whereas with augmentation the accuracy increased to $97.2\%$. We have tested the generalization of our method with five unseen WSIs from the NIH (National Cancer Institute) dataset. The accuracy for patch-wise detection was $81\%$. One folded patch within an image suffices to flag the entire specimen for visual inspection.

\end{abstract}
\section{Introduction}

For most types of cancer, biopsy is a dominant procedure for diagnosis. During the biopsy, a small part of suspicious tissue is cut out. After tissue preparation, a tiny section of tissue is mounted on a glass slide. Pathologists visually inspect these glass slides under a microscope and write a report to justify a primary diagnosis \cite{zerbino1994biopsy}. 

The rapid progress of image acquisition technologies over the past decade has led to a dramatic change in the pathology field by developing digital pathology. Most whole slide scanners can produce a high-resolution digital image of histology glass slides in a few minutes \cite{al2012digital}. These WSIs can be analyzed on a display rather than through the microscope. In addition, sharing scans for teleconsultation purposes are much more convenient in digital version compared to shipping the glass slides to other laboratories to solicit a second opinion \cite{tizhoosh2018artificial}. 

\begin{figure}[t]
\centering 
\stackunder[5pt]{\includegraphics[width=0.28\columnwidth,height=0.27\columnwidth]{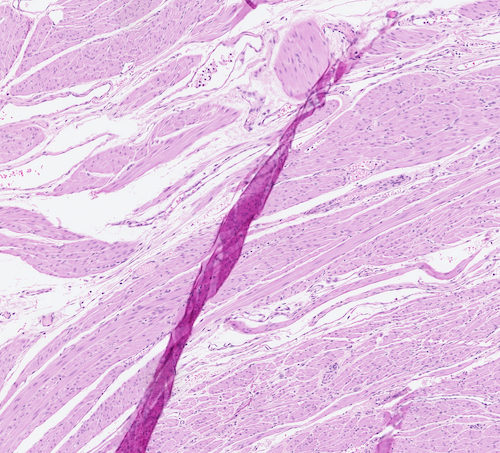}}{\tiny }
\stackunder[5pt]{\includegraphics[width=0.28\columnwidth,height=0.27\columnwidth]{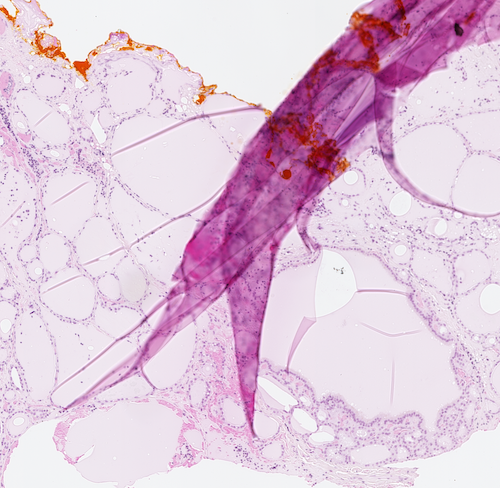}}{\tiny }
\stackunder[5pt]{\includegraphics[width=0.28\columnwidth,height=0.27\columnwidth]{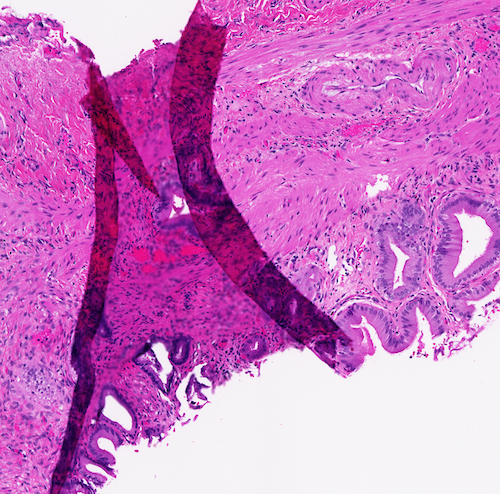}}{\tiny }
\caption{Sample folded tissues from our dataset.}
\label{fig:Sample}
\end{figure}

Regardless whether we use digitization or microscopy, the presence of artifacts such as folded tissue might negatively affect the diagnosis \cite{bindhu2013facts}. When digital technology is used other artifacts like blur  may also reduce the quality of computerized algorithms \cite{kothari2012biological}. Tissue fold can occur in the sectioning part of tissue processing when a thin tissue slice is folded \cite{kothari2013eliminating}. Figure \ref{fig:Sample} shows three samples of folded tissue. 

The difference in tissue thickness changes the precise lens focus when one or more focus points are localized on the folded parts. Most WSI scanners start with a lower resolution pre-scan phase which selects focus points in the some areas with possibility of manual adjustment. Tissue-fold detection can avoid placing  focus points on the folded areas. In addition, different tissue cuts are available in the laboratories. Selecting a suitable glass slide by a rapid pre-scan quality control system could save valuable scanning time and improve the workflow. 

\section{Related Works}
The research on tissue-fold detection is relatively young. Pinky et al. \cite{bautista2009detection} proposed a technique to use colour information to detect tissue folds. The fact that the saturation of the folded area is different from other parts justifies the development of a colour shifting method to magnify the colour metric difference in folded and non-folded areas  \cite{bautista2010improving}. Other authors have suggested adding the intensity level to the saturation criteria to develop a fold segmentation method. In such algorithms, if saturation minus intensity is higher than a certain threshold, this area is segmented as a folded tissue  \cite{kothari2012biological}. More recently, statistical approaches such as the rank-sum method have been applied to find image features (e.g., colour and connectivity descriptors) that are discriminated from the same set of WSIs with and without folds \cite{kothari2013eliminating}.

Generally, there is an inevitable  drawback associated with the use of colour information as a feature for tissue-fold distinction. Colour-based approaches might easily fail due to the colour fluctuations which occur in digital pathology relatively often. These changes might happen mainly because of  ``differences in manufacturing techniques of stains, laboratories' staining protocols, and colour responses of digital scanners'' \cite{vahadane2016structure}. Convolutional neural networks (CNNs), on the other hand, have been widely used recently in almost every field of machine vision due to their unique ability to capture accurate data-driven features \cite{khatami2018parallel}. As result of augmentation techniques in their training process, these networks are fairly robust to a variety of changes including colour changes \cite{lafarge2017domain}. As a matter of fact most CNNs are trained through diverse augmentation techniques, among others variations of color. 
In the deep learning literature, the importance of labeled data is undeniable. Deep networks need more and more labeled training data to train each layer's parameters when the networks become deeper and deeper \cite{szegedy2015going}. On the other hand, providing a large number of labeled data in the medical domain by the expert physicians is expensive. In contrast, transfer learning is considered to be an applicable solution to fine-tune a pre-trained deep network with a much smaller training set compared to training from scratch. The idea behind transfer learning is that if the network is trained with a large dataset such as ImageNet \cite{deng2009imagenet}, it learns useful (general) information that can be applied in completely different domains. In general, for any given pre-trained network, the first layers will be held unchanged (i.e., \emph{frozen}), while the weights of the last few layers will be adjusted by re-training with the data of the new domain \cite{pan2010survey,kumar2018deep}. Moreover, using deep pooling or the weights of fully connected layers have been reported to be excellent sources for feature extraction \cite{kieffer2017convolutional,van2015off}.

In this work, we have compared five well-known pre-trained CNNs as feature extractors for classifying folded tissue against normal tissue. VGG16, GoogleNet, Inception V3, ResNet 101 and DenseNet 201 are the networks that have  compared in our experiments. We also examined the discrimination power of decision trees, SVM and k-NNs with respect to the classification of different deep features. 

\section{Materials and Methods}
\subsection{Folded-Tissue Dataset}
In our experiments, we created a training dataset of folded tissue images. We had access to 79 \emph{rejected} WSIs from Huron Digital Pathology \footnote{http://www.hurondigitalpathology.com/}. These scans had been rejected due to presence of different artifacts. Since there was a large number of folded-tissue cases in these slides, we created a folded-tissue training dataset and did not consider other types of artifacts. The folded regions are selected on fairly large windows at $20$x magnification (about $5000 \times 5000$ pixels) which is roughly equivalent to $250 \times 250$ pixels at $1$x magnification. In practice, low magnification images could be easily obtained in the fast pre-scan mode.   

We have manually selected 112 folded-tissue patches as the training set through visual inspection. Since we needed to classify them against the normal (unfolded) tissue, we selected 315 images from the area around the folded regions as negative samples (i.e., unfolded tissue). We augmented each image to 12 images by rotating ($0^\circ$ and $90^\circ$),  flipping (flipped/no-flipped) and changing the illumination  (original, suppressed, amplified). As a result, we established a dataset of 1,344 folded patches and 3,780 fold-free patches. Change in illumination was done by converting each image to the CIELAB colour space (LAB), amplifying and suppressing the L channel by a factor of 1.25 and 0.75, respectively. Finally, we converted them back to RGB colour space. At the end, all patches were re-sized to input size of each pre-trained network required size (e.g. $255 \times 255$ for DenseNet).    

To evaluate the generalization and the practical performance, we also selected five WSIs from the NIH dataset\footnote{https://gdc.cancer.gov}. We selected WSIs from three different organs (kidney,lung and colon). Figure \ref{fig:WSI samples} shows two sample WSIs alongside the boundary boxes of our classifier. 

\subsection{Pre-Trained CNNs}
\label{S:cnn}
CNNs are a class of deep networks designed to learn a large bank of filters. These filters are convolved with the input image in a hierarchical fashion. The major advantage of CNNs is their independence from prior knowledge and handcrafted feature design. There are several deep networks that have been trained with available public images and can be employed for classification in different domains. VGG16 \cite{simonyan2014very}, GoogleNet  \cite{szegedy2015going}, Inception-V3 \cite{szegedy2016rethinking}, ResNet \cite{he2016deep} and DenseNet-201 are major examples for pre-trained networks. 


DenseNet-201 is a CNN that is designed to  overcome the \emph{gradient vanishing} problem by adding dense blocks and transition layers. The vanishing gradient prevents a network  from growing. As a result of DenseNet extensions, the network learns rich feature representations for a wide range of images due to its extremely deep architecture. We used the last fully connected layer of the network with 1024 elements as the feature vector. 

\section{Experiments and Results}
We experimented with several learning methods to classify the dense features including SVM \cite{suykens1999least}, decision trees \cite{safavian1991survey} and k-NNs \cite{xu2013coarse} to find the optimal classifier for tissue fold detection. As listed in Table \ref{table:classifiers}, the ability of the quadratic SVM  to classify the folded and non-folded tissue was the highest with 96.3\% accuracy. However, median Gaussian SVM and fine k-NN also achieved acceptable results with accuracy values of 94.8\% and 94.1\%, respectively.
 Table \ref{table:classifiers} compares the performance of different classifiers when DenseNet features were used. 

Augmentation and leave-one-out schemes were selected to compensate for the small size of training data. Since the size of the training data was relatively small, we applied augmentation techniques to increase the number of observations. As well, the leave-one-out strategy \cite{kohavi1995study} to evaluate the accuracy was used to perform as many experiments as possible.          

\begin{table}[h]
\begin{minipage}[b]{70mm}
\begin{tabular}{l|c|c|c|c|c|}
\textit{\textbf{}} & \cellcolor[HTML]{EFEFEF}Network & \multicolumn{1}{l|}{\cellcolor[HTML]{EFEFEF}Depth} & \cellcolor[HTML]{EFEFEF}Sensitivity & \cellcolor[HTML]{EFEFEF}Recall  & \multicolumn{1}{l|}{\cellcolor[HTML]{EFEFEF}accuracy} \\ \cline{2-6} 
1 & VGG-16 & 16 & 87.55\% & 90.8\% & 91.8\% \\
2 & Google Net & 22 & 90.8\% & 92.6\% & 93.7\% \\
3 & inceptionv3 & 48 & 90.5\% & 92.85\% & 93.7\% \\
4 & resnet101 & 101 & 93.2\% & 92.95\% & 94.6\%  \\
5 & densenet201 & 201 & \cellcolor[HTML]{ECF4FF}94.6\%  & \cellcolor[HTML]{ECF4FF}96.85\% & \cellcolor[HTML]{ECF4FF}96.7\%                
\end{tabular}
\vspace{1mm}
\caption{The accuracy of five pre-trained networks.}
\label{table:cnns}
\end{minipage}
\begin{minipage}[b]{60mm}
\hspace{.5cm}
\begin{tabular}{l|c|c|c|}
\cline{2-4}
  & \cellcolor[HTML]{EFEFEF}Classifier & \cellcolor[HTML]{EFEFEF}Sub-type & \cellcolor[HTML]{EFEFEF}accuracy \\ \cline{2-4} 
1 & Tree & Fine & 81.5\%\\
2 & Tree & Coarse & 82.9\% \\
3 & SVM & Quadratic & \cellcolor[HTML]{ECF4FF}96.3\%   \\
4 & SVM & Med Gaussian & 94.8\% \\
5 & k-NN & Fine & 94.1\% \\
6 & k-NN & Cosine & 87.1\% \\ \cline{2-4} 
\end{tabular}
\vspace{1mm}
\caption{The accuracy of different classifiers.}
\label{table:classifiers}
\end{minipage}
 \end{table}
 
\begin{figure}[h]
\centering 
\stackunder[5pt]{\includegraphics[width=50mm,height=50mm]{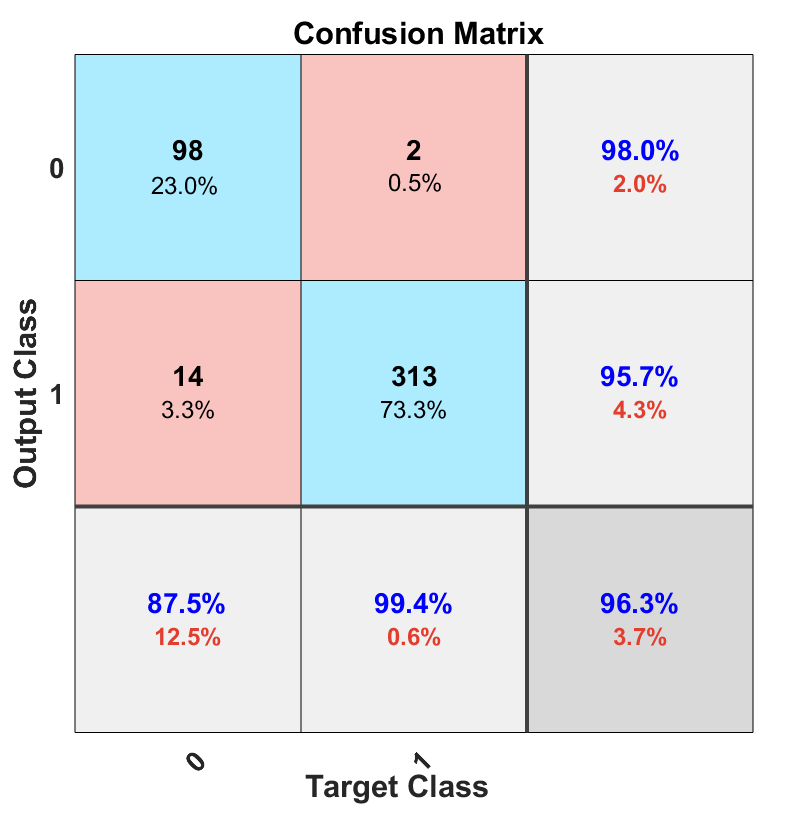}}{\tiny }
\stackunder[5pt]{\includegraphics[width=50mm,height=50mm]{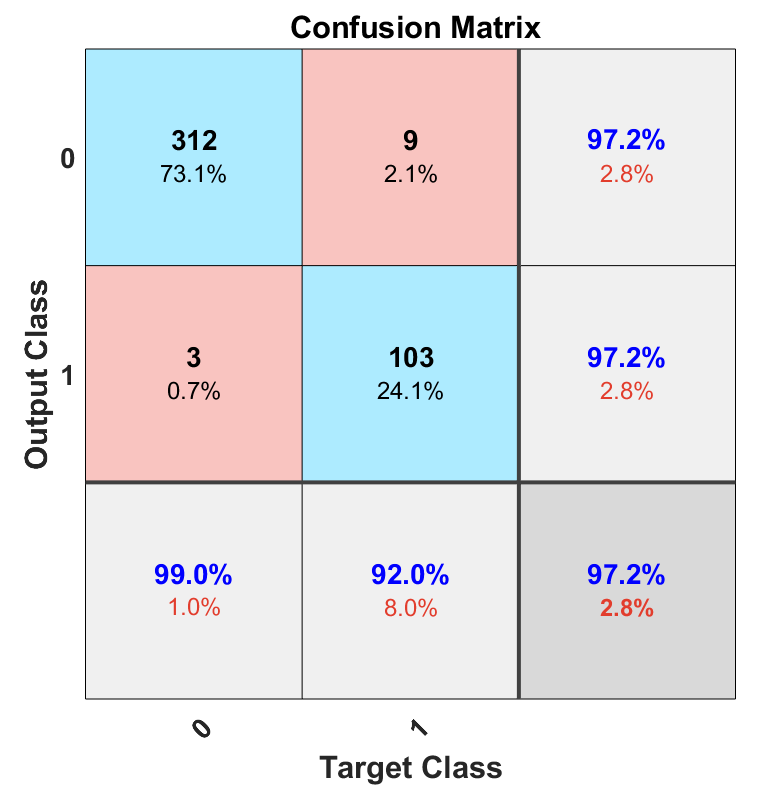}}{\tiny }
\caption{Confusion matrices of folded tissue classification. The left matrix depicts the classification without augmentation while the right matrix shows the values after  augmentation.}
\label{fig:confusion}
\end{figure}

Fig. \ref{fig:confusion} shows the confusion matrices of leave-one-out quadratic SVM with augmentation on the right side versus no augmentation on the left side. 
By applying the augmentation method, not only did the total accuracy increase slightly to 97.2\% but also the false negative (folded patch, but classified as normal patch) also decreased from 2.6\%  to 2.1\%. However, the false positive (normal patch, but classified as a folded patch) remained unchanged. In general, any type of error is not desirable, nevertheless, in our case, false positive might be preferred over false negative. 

Table \ref{table:cnns} shows the performance of the networks  when their features were classified by SVM. As it can be seen from the table, the performance of deep features increase in our application when the depth of network increased.  

\begin{figure*}[h]
\centering
\includegraphics[width=55mm]{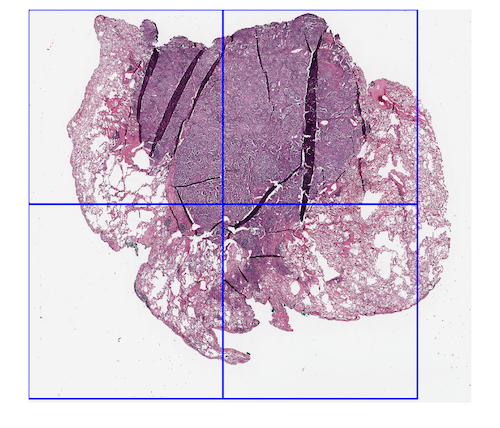}
\includegraphics[width=55mm]{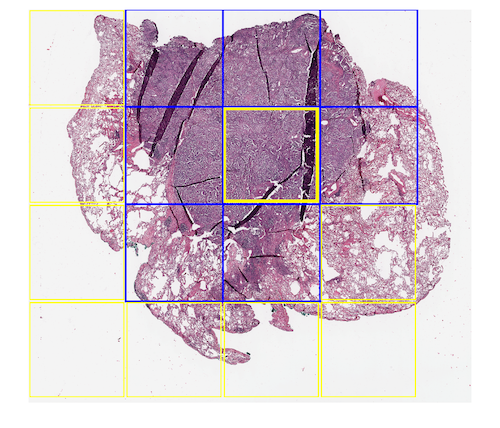}
\includegraphics[width=110mm]{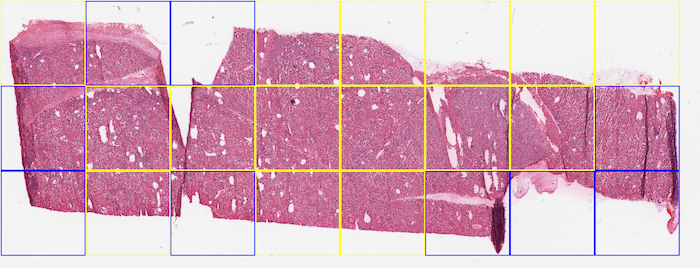}
\caption{Results of applying our classifier for two selected WSIs from NIH dataset (Lung and kidney) - Folded classes are distinguished by the blue boxes while the yellow colour is used for normal tissue: (top left)  $3000\times3000$ patches are fed to the classifier, (top right) $1500\times1500$ patches are fed to the classifier, and (bottom) a large scan with  $4000\times4000$ patches. }
\label{fig:WSI samples}
\end{figure*}

Classifiers, which are trained on small datasets predominantly fail to generalize on new classification categories. In our experiments, we selected five new WSIs with a noticeable amount of folded tissue from the NIH dataset to evaluate the ability of deep features and SVM to generalize to unseen cases. We applied our method in different window sizes with no overlap. All patches will be resized to $255 \times 255$ before feeding to the network. Figure \ref{fig:WSI samples} shows sample WSIs from NIH database with different window sizes. Blue boxes are representative of the presence of folded tissue while a yellow box represents normal tissue. The overall accuracy in generalization test set with $4000 \times 4000$ pixel size dropped to $81\%$. A possible explanation for this result may be the lack of adequate fold pattern samples in the training set. Besides the difference in an organ type, scanner brand should also be considered. However, as we trained and tested the classifiers for patch-wise tissue detection, one has to bear in mind that the detection of one tissue fold is sufficient to flag a scan for visual inspection. 

It can be seen in Figure \ref{fig:WSI samples}(b) that there is a folded patch which has not been detected. There might be some justifications for this false negative -yellow window in Figure \ref{fig:WSI samples} (b)-. The first one is that our training dataset enclosed the entire folded tissue within each patch (i.e., no folded tissue was split between two patches). In this false negative example, however, the patch does not contain all of the folded tissue, and parts of the folded tissue are contained within the neighbouring patches. The same error has occurred in Figure \ref{fig:WSI samples}(c). 
The second justification is that the training patch sizes were about 5000 by 5000 pixels, while the experiment window size was 1500 by 1500, therefore training with bigger size patches (patch in lower magnification) might have been the reason for false negatives.    

\section{Conclusions}
Quality control for artifact detection in histopathology slides could be used in order to reject defective slides. This procedure may save time in clinical practice. Not only can folded tissue, as one of the most common artifacts in histopathology slides, lead to rejection of slides in clinical practice, but it may also negatively affect the diagnosis. In this paper, a procedure based on deep features has been proposed to detect folded tissues in large scan regions. We trained an SVM classifier based on the features of augmented training patches to classify folded and normal tissues. The accuracy in the presented dataset was quite high, whereas the model's generalization on new WSIs was acceptable.

Several topics can be anticipated for the future works. A larger dataset, with a known source of organ could boost up the generalization. And different patch size selection for dataset also could help to boost the accuracy.

\bibliographystyle{splncs04}
 \bibliography{Ref.bib}
\end{document}